\documentclass[english,twocolumn,prl,superscriptaddress,nofootinbib,preprintnumbers]{revtex4}
\usepackage[T1]{fontenc}
\usepackage[latin1]{inputenc}
\usepackage{amsmath}
\usepackage{graphicx}
\usepackage{babel}

\makeatletter

\usepackage{dcolumn}

\makeatletter

\usepackage{ifpdf}
\usepackage{epstopdf}
\graphicspath{{./}}

\newcommand{\dd}{\textrm{d}}

\makeatother

\makeatother

\begin{document}

\title{Cosmic Parallax: possibility of detecting anisotropic
expansion of the universe by very accurate astrometry measurements}

\author{Claudia Quercellini}
\affiliation{Università di Roma Tor Vergata, Via della Ricerca Scientifica 1,
00133 Roma, Italy}

\author{Miguel Quartin}
\affiliation{INAF/Osservatorio Astronomico di Roma, V. Frascati 33, 00040 Monteporzio
Catone, Roma, Italy}
\affiliation{Universitità di Milano-Bicocca, Dip. Fisica {}``G. Occhialini'',
P.le Scienze 3, 20126 Milano, Italy}

\author{Luca Amendola}
\affiliation{INAF/Osservatorio Astronomico di Roma, V. Frascati 33, 00040 Monteporzio
Catone, Roma, Italy}

\date{\today{}}

\begin{abstract}
Refined astrometry measurements allow us to detect large-scale deviations from isotropy through real-time observations of changes in the angular separation between sources at cosmic distances. This ``cosmic parallax'' effect is a powerful consistency test of FRW metric and may set independent constraints on cosmic anisotropy. We apply this novel general test to LTB cosmologies with off-center observers and show that future satellite missions such as Gaia might achieve accuracies that would put limits on the off-center distance which are competitive with CMB dipole constraints.
\end{abstract}

\maketitle

\label{sec:intro} \emph{Introduction}. The standard model of cosmology
rests on two main assumptions: general relativity and a homogeneous
and isotropic metric, the Friedmann-Robertson-Walker metric (henceforth
FRW). 
While general relativity has been tested with great precision at least
in laboratory and in the solar system, the issue of large-scale deviations
from homogeneity and isotropy is much less settled. There is by now
an abundant literature on tests of the FRW metric, and on alternative
models invoked to explain the accelerated expansion by the effect
of strong, large-scale deviations from homogeneity (see
\cite{Celerier:2007} for a review). Several such models adopt as an alternative
the Lemaître-Tolman-Bondi metric (henceforth LTB), that is a model
with a spherically symmetric distribution of matter (see for instance~\cite{Marra:2007,Notari:2007,Bassett:2008,Alnes:2006a,Bellido:2008,2008arXiv0807.1443C,Humphreys:1997}).
The main motivation for this is the fact that the distance-dependent
expansion rate can explain the supernovae Ia excess dimming without
a dark energy field.

LTB universes appear anisotropic to any observer except the central one. In every anisotropic expansion the angular separation between any two sources varies in time, thereby inducing a \emph{cosmic parallax} (CP) effect. This is totally analogous to the classical stellar parallax, except here the parallax is induced by a differential cosmic expansion rather than by the observer's own movement. Together with the Sandage effect of velocity shift $\dot{z}$ \cite{Sandage:1962,loeb98}, CP belongs to the new realm of \emph{real-time cosmology}, a direct way of testing our universe based on cosmological observations spaced by several years~\cite{Lake:2007,Balbi:2007,Corasaniti:2007,Uzan:2008,Uzan:2008b}.

One can expect on dimensional grounds that this differential cosmic
expansion generates after an observation time lag of $\Delta t=10$~years
an effect of order $H_{0}\Delta t=10^{-9}h$; therefore two sources
separated by 1~rad today will show a parallax of $10^{-9}h~\mbox{rad}\approx200h~\mu\mbox{as}\,$
in ten years, which is well above the accuracy goal of $10~\mu\mbox{as}\,$
set by Gaia (for V$\le$15)~\cite{Bailer:2004} and other planned
missions like SIM \cite{Goullioud:2008}, JASMINE~\cite{Yano:2008} and VSOP-2 \cite{2007IAUS..242..517M}.
Adopting Gaia specifications, we show that $10^{6}$ quasars might be enough to constrain the off-center distance $r_{0}$ to the $10$ Megaparsec scale, similar to or better than any other proposed test of the Copernican principle but without the degeneracy with our peculiar velocity that afflicts the constraints from the cosmic microwave background~\cite{Alnes:2006a,Humphreys:1997,Caldwell:2008}. 
In this paper we first derive an estimate of CP by intuitive arguments based on a FRW description and then test it with a full numerical integration of light-ray geodesics in LTB models.

\begin{figure}[t!]
\includegraphics[width=5.5cm]{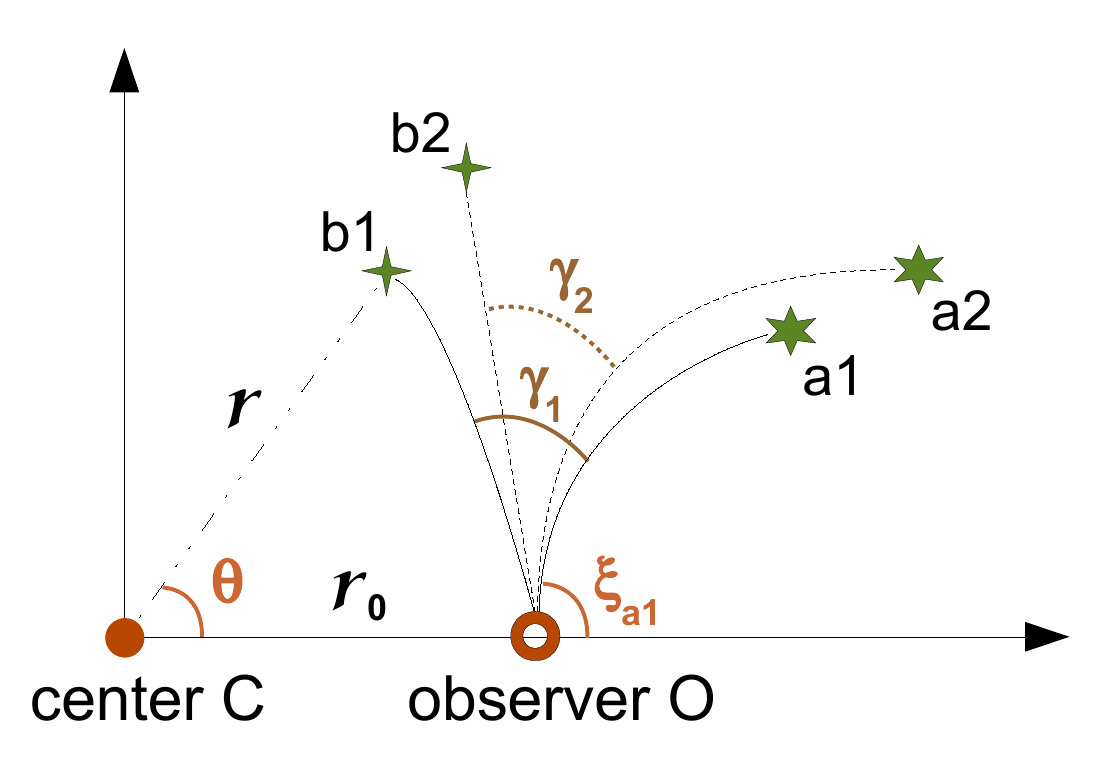} \vspace{-0.25cm}
\caption{Overview, notation and conventions. For clarity purposes
we assumed here that the points $C,O,a_{1},b_{1},a_{2},b_{2}$ all lie on the
same plane. Comoving coordinates $r$ and $r_{0}$ correspond to physical coordinates $X$ and $X_{0}$.}
\vspace{-0.3cm}
\label{fig:overview}
\end{figure}

\emph{Estimating the cosmic parallax.} Figure~\ref{fig:overview} depicts
the overall scheme describing a possible time-variation of the angular
position of a pair of sources that expand anisotropically with respect
to the observer. We label the two sources $a$ and $b$, and the two
observation times 1 and 2. In what follows, we will refer to ($t$,
$r$, $\theta$, $\phi$) as the comoving coordinates with origin
on the center of a spherically symmetric model. Peculiar velocities
apart, the symmetry of such a model forces objects to expand radially
outwards, keeping $r$, $\theta$ and $\phi$ constant.

Let us assume now an expansion in a flat FRW space from a {}``center''
$C$ observed by an off-center observer $O$ at a distance $X_{0}$
from $C$. Since we are assuming FRW it is clear that any point in
space could be considered a {}``center'' of expansion: it is only
when we will consider a LTB universe that the center acquires an absolute
meaning. The relation between the observer line-of-sight angle $\xi$
and the coordinates of a source located at a radial distance $X$
and angle $\theta$ in the $C$-frame is \vspace{-0.2cm}
 \begin{equation}
\cos\xi=\frac{X\cos\theta-X_{0}}{(X^{2}+X_{0}^{2}-2\, X_{0}X\cos\theta)^{1/2}},\vspace{-.2cm}\end{equation}
 where all angles are measured with respect to the $CO$ axis and
all distances in this section are to be understood as physical distances.

We consider first two sources at location $a_{1},b_{1}$ on the same
plane that includes the $CO$ axis with an angular separation $\gamma_{1}$
as seen from $O$, both at distance $X$ from $C$ (throughout this
letter we shall always assume for simplicity that both sources share
the same $\phi$ coordinate). After some time $\Delta t$, the sources
move to positions $a_{2},b_{2}$ and the distances $X$ and $X_{0}$
will have increased by $\Delta_{t}X$ and $\Delta_{t}X_{0}$ respectively.
If for a moment we allow ourselves
the liberty of assigning to the scale factor $a(t)$ and the $H$
function a spatial dependence, a time-variation of $\gamma$ is induced.
The variation $\Delta_{t}\gamma$ is the anticipated cosmic parallax
effect and can be easily estimated if we suppose that the Hubble law
is just generalized to $\Delta_{t}X=XH(t_{0},X)\Delta t\equiv XH_{X}\Delta t$.
Generalizing to sources on different shells separated by a small $\Delta X\equiv X_{b}-X_{a}$
(not to be mistaken with the \emph{time} interval $\Delta_{t}X$)
and $\Delta\theta\equiv\theta_{b1}-\theta_{a1}$, after straightforward
geometry we arrive at\vspace{-0.15cm}
\begin{equation}
\begin{aligned}\Delta_{t}\gamma\,=\, & s\,\Delta t(H_{{\rm obs}}-H_{X})\left(\cos\theta\,\Delta\theta+\sin\theta\,\Delta X/X\right)\\
 & +H_{X}\dd H_{X}/\dd X\,\sin\theta\,\Delta X+O(s^{2})\,.
\end{aligned}
\vspace{-.1cm}\label{eq:deltagamma}
\end{equation}
where $H_{{\rm obs}}\equiv H(t_{0},r_{0})$, $s\equiv X_{0}/X\ll1$
(at this order we can neglect the difference between the observed
angle $\xi$ and $\theta$). We can also convert the above intervals
$\Delta X$ into the redshift interval $\Delta z$ by using the relations
$r=\int\dd z/H(z)$ and $X=\int a(t_{0},r)\dd r$, which combine to
$\Delta X=a(t_{0},X)\Delta z/H(z)\sim\Delta z/H(z)$ (we impose $a(t_{0},X_{0})=1$),
where $H(z)\equiv H(t(z),X)$.

In a FRW metric, $H$ does not depend on $r$ and the parallax vanishes.
On the other hand, any deviation from FRW entails such spatial dependence
and the emergence of cosmic parallax, except possibly for special
observers (such as the center of LTB). A constraint on $\Delta_{t}\gamma$
is therefore a constraint on cosmic anisotropy.

Rigorously, one actually needs to perform a full integration
of light-ray geodesics in the new metric. Nevertheless, we shall assume
for a moment that for an order of magnitude estimate we can simply
replace $H$ with its space-dependent counterpart given by LTB models.
In order for a LTB cosmology to have any substantial
effect (e.g., explaining the SNIa Hubble diagram) it is reasonable
to assume a difference between the local $H_{{\rm obs}}$ and the
distant $H_{X}$ of order $H_{{\rm obs}}$~\cite{Alnes:2006a}.
More precisely, putting $H_{obs}-H_{X}=H_{obs}\Delta h$ then one
has that the overall effect $\Delta_{t}\gamma$ for on-shell sources
is of order $\, s\cos\theta H_{{\rm obs}}\,\Delta h\Delta t\Delta\theta$. That is, as anticipated, for $\,\Delta t=10$~years we expect a
parallax of $\,200s\cos{\theta}\Delta h\Delta\theta\,\mu$as for sources separated
by $\Delta\theta$. Similarly, for radial source pairs, neglecting
the $dH_{X}/dX$ term (which is valid except close to the edge of
the LTB void), one has $s\sin{\theta}H_{{\rm obs}}\,\Delta h\Delta t\,\Delta z/z\approx200s\sin{\theta}\,\Delta h\Delta z/z\,\,\mu$as
(assuming $X\approx zH(z)^{-1}$).

Let us finally consider the main expected source of noise, the intrinsic
peculiar velocities of the sources. The variation in angular separation
for sources at angular diameter distance $D_{A}$ (measured by the
observer) and peculiar velocity $v_{{\rm pec}}$ can be estimated
as \vspace{-0.2cm}
\begin{equation}
    \Delta_{t}\gamma_{{\rm pec}}=\left(\frac{v_{{\rm pec}}}{500\,\mbox{km/s}}\right) \left(\frac{D_{A}}{1\,\mbox{Gpc}}\right)^{\!-1}\!\! \left(\frac{\Delta t}{10\,\mbox{years}}\right)\mu{\normalcolor as}.\vspace{-.1cm}\label{eq:pecvel}
\end{equation}
This velocity field noise is therefore typically smaller than the
experimental uncertainty (especially for large distances) and again
will be averaged out for many sources. Notice that the observer's
own peculiar velocity produces a systematic offset sinusoidal signal
$\Delta_{t}\gamma_{{\rm pec},O}$ of the same amplitude as $\Delta_{t}\gamma_{{\rm pec}}$
that has to be subtracted from the observations: we discuss this further
below.

\emph{Geodesic Equations}. \label{sec:geodesic} These very suggestive
but simplistic calculations need confirmation from an exact treatment
where the full relativistic propagation of light rays is taken into
account. We will thus consider in what follows two particular LTB
models capable of fitting the observed SNIa Hubble diagram and the
CMB first peak position and compatible with the COBE results of the
CMB dipole anisotropy, as long as the observer is within around 15
Mpc from the center~\cite{Alnes:2006a}. Moreover, both models have
void sizes which are small enough ($z\sim0.3$) not to be ruled out
due to distortions of the CMB blackbody radiation spectrum~\cite{Caldwell:2008}.

The LTB metric can be written as (primes and dots refer to partial
space and time derivatives, respectively):\vspace{-0.15cm}
 \begin{equation}
{\textrm{d}}s^{2}=-{\textrm{d}}t^{2}+\frac{\left[R'(t,r)\right]^{2}}{1+\beta(r)}{\textrm{d}}r^{2}+R^{2}(t,r){\textrm{d}}\Omega^{2},\vspace{-.15cm}\label{eq:LTB}\end{equation}
 where $\beta(r)$ can be loosely thought as position dependent spatial
curvature term. Two distinct Hubble parameters corresponding to the
radial and perpendicular directions of expansion are defined as $H_{||}=\dot{R'}/R'$
and $H_{\perp}=\dot{R}/R$ (in a FRW metric $R=ra(t)$
and $H_{||}=H_{\perp}$). This class of models exhibits implicit analytic
solutions of the Einstein equations in the case of a matter-dominated
universe, to wit (in terms of a parameter $\eta$) \vspace{-0.2cm}
\begin{align}
    R=\, & (\cosh\eta-1)\frac{\alpha}{2\beta}+R_{{\rm lss}}\left[\cosh\eta+\sqrt{D}\sinh\eta\right],\vspace{-.1cm}\label{eq:sol-R}\\
    \sqrt{\beta}t=\, & (\sinh\eta-\eta)\;\alpha/(2\beta)\,+\nonumber \\
    & +R_{{\rm lss}}\left[\sinh\eta+\sqrt{D}\left(\cosh\eta-1\right)\right],\vspace{-.2cm}\label{eq:sol-beta-t}
\end{align}

\vspace{-.2cm}
\noindent where $D=(\alpha+\beta R_{{\rm lss}})/(\beta R_{{\rm lss}})$, and
$\alpha$, $\beta$ and $R_{{\rm lss}}$ are all functions of $r$.
In fact, $R_{{\rm lss}}(r)$ stands for $R(0,r)$ and we will choose
$t=0$ to correspond to the time of last scattering, while $\alpha(r)$
is an arbitrary function and $\beta(r)$ is assumed to be positive.
Due to the axial symmetry and the fact that photons follow a path
which preserves the 4-velocity identity $u^{\alpha}u_{\alpha}=0$,
the four second-order geodesic equations for $(t,r,\theta,\phi)$
can be written as five first-order ones. We will choose as variables
the center-based coordinates $t$, $r$, $\theta$, $p\equiv\dd r/\dd\lambda$
and the redshift $z$, where $\lambda$ is the affine parameter of
the geodesics. We shall refer also to the conserved angular momentum $J\equiv R^{2}\dd\theta/\dd\lambda=\textrm{\emph{const}}=J_{0}$.
For a particular source, the angle $\xi$ is the coordinate equivalent
to $\theta$ for the observer, and in particular $\xi_{0}$ is the
coordinate $\xi$ of a photon that arrives at the observer at the
time of observation $t_{0}$. Obviously this coincides with the measured
position in the sky of such a source at $t_{0}$. In terms of these
variables, and defining $\lambda$ such that $u(\lambda)<0$, the
autonomous system governing the geodesics is written as \vspace{-0.2cm}
\begin{equation}\label{eq:geodesics}
\begin{aligned}
    \frac{\dd t}{\dd\lambda}\,=\; & -\sqrt{\frac{(R')^{2}}{1+\beta}\, p^{2}+\frac{J^{2}}{R^{2}}}\,,\;\;\frac{{\textrm{d}}r}{{\textrm{d}}\lambda}\,=\, p\,,\;\;\frac{\dd\theta}{\dd\lambda}\,=\,\frac{J}{R^{2}}\,,\hspace{-3cm}\\
    \frac{\dd z}{\dd\lambda}\,=\; & \frac{(1+z)}{\sqrt{\frac{(R')^{2}}{1+\beta}\, p^{2}+\frac{J^{2}}{R^{2}}}}\left[\frac{R'\dot{R}'}{1+\beta}\, p^{2}+\frac{\dot{R}}{R^{3}}J^{2}\right],\\
    \frac{\dd p}{\dd\lambda}\,=\; & 2\dot{R}'\, p\,\sqrt{\frac{p^{2}}{1+\beta}\,+\frac{J^{2}}{R^{2}\, R'^{2}}}\,+\,\frac{1+\beta}{R^{3}R'}J^{2}\,+\\
     & +\left[\frac{\beta'}{2+2\beta}-\frac{R''}{R'}\right]p^{2}\,.
\end{aligned}
\end{equation}
\vspace{-.2cm}

Following~\cite{Alnes:2006a}, the angle $\xi$ along a geodesic is given by $\cos\xi = -R'(t,r)\, p\,/\,(u\,\sqrt{1+\beta(r)})$, from which we obtain $p_{0}\,=\,-\sqrt{1+\beta(r_{0})}\cos(\xi_{0})\,/\, R'(t_{0},r_{0})$ and $J_0= J=  R(t_{0},r_{0})\sin(\xi_{0})$. Therefore, our system is completely defined by the initial conditions $t_{0}$, $r_{0}$, $\theta_{0}=0$, $z_{0}=0$ and $\xi_{0}$. The first two define the instant of measurement and the offset between observer and center, while $\xi_{0}$ stands for the direction of incidence of the photons. By integrating the geodesic equations for two sources located at $(z_{a1},z_{b1},\xi_{a1},\xi_{b1})$ after a time interval $\Delta t$ the CP will be
$\Delta_{t}\gamma\equiv\gamma_{2}-\gamma_{1}=(\xi_{a2}-\xi_{b2})-(\xi_{a1}-\xi_{b1})$.

The models of Ref.~\cite{Alnes:2006a} are characterized by a smooth transition between an inner void and an outer region with higher matter density and are described by the functions $\alpha(r)$ and $\beta (r)$ (Eqs. (28) and (29) in~\cite{Alnes:2006a}), themselves carrying a total of 4 free parameters, one of which the value $H_{\perp,0}^{{\rm out}}$ of the Hubble constant at the outer region, set at 51~km/(s$\,\mbox{Mpc}$). Following~\cite{Alnes:2006a} we dub them Model I and II; the main difference between them is that Model II features a sharper transition from the void. However transition width is not expected to be an important factor in CP since most quasars are outside the void and the most relevant quantity is the difference between the inner and outer values of $H$. In both cases we set the off-center (physical) distance to 15~Mpc, which is the upper limit allowed by CMB dipole distortions~\cite{Alnes:2006a},\linebreak and this corresponds to $s\simeq4.4\;10^{-3}$ for a source at $z=1$.  It can be shown that in a step-like LTB void model, the $H_{X}$ in~\eqref{eq:deltagamma} is given by $H_{||,0}^{{\rm in}}\, X_{{\rm vo}}/X + H_{||,0}^{{\rm out}}\,(1-X_{{\rm vo}}/X)\approx H_{||,0}^{{\rm out}}$, whereas $H_{{\rm obs}}\approx H_{||,0}^{{\rm in}}\,$.

\begin{figure}[t!]
\includegraphics[width=6.5cm]{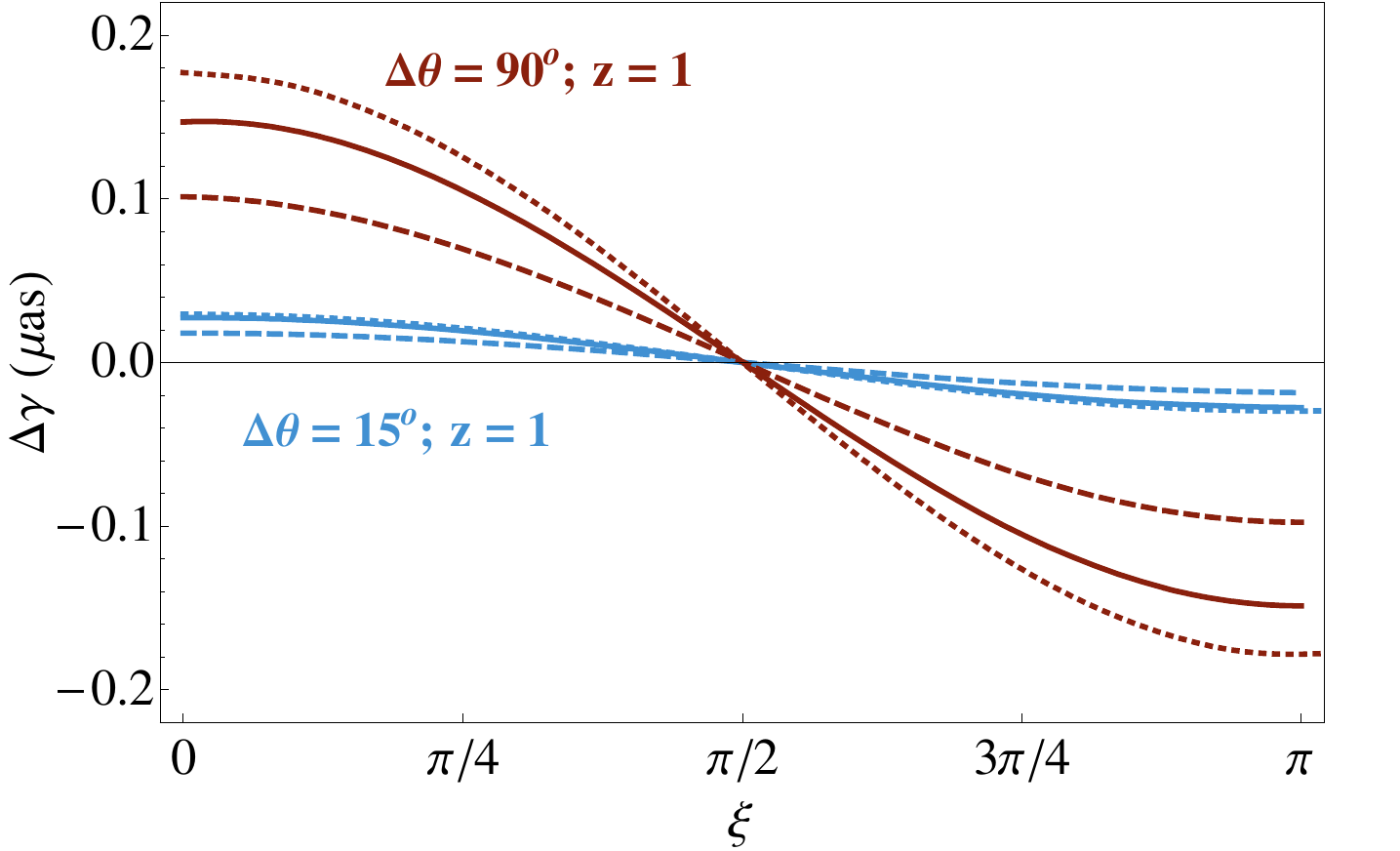}
\vspace{-0.2cm}
\caption{$\Delta_{t}\gamma$ for three sources at the same shell, at $z=1$,
for both models I (full lines) and II (dashed), and the FRW-like estimate
(dotted). 
As
expected, $\Delta_{t}\gamma$ is linear in $\Delta\theta$.}
\vspace{-0.2cm}
 \label{fig:dgammath}
\end{figure}

In Figure~\ref{fig:dgammath} we plot $\Delta_{t}\gamma$ for three
sources at $z=1$, for models I and II as well as for the FRW-like
estimate. One can see that the results do not depend sensitively on
the details of the shell transition and that in both cases the FRW-like
estimate gives a reasonable idea of the true LTB behavior. We conclude
that~\eqref{eq:deltagamma} is a valid approximation. Numerically,
we find that a convenient estimate of the parallax is given by $\Delta_{t}\gamma=\,100\: s\Delta h(\Delta t/10\mbox{ years})\times(\Delta\theta,\Delta z/z)\mu$as. For the LTB models above
$\Delta h\approx0.12-0.14$.

Integrating~\eqref{eq:geodesics} yields also another interesting observable, $\dot{z}=\Delta_{t}z/\Delta t$,
coupled to the CP: together they form a new set of real-time
cosmic observables. Ref.~\cite{Uzan:2008} calculated $\dot{z}$
for an observer at the center of a LTB model. In the limit of small
$s$ our numerical results reveal that $\dot{z}$ for off-center observers
show very small angular dependence.

\emph{Cosmic parallax with Gaia.} A realistic possibility of observing
the CP is offered by the forthcoming Gaia mission. Gaia
will produce in five years a full-sky map of roughly $500,000$ quasars; by making of order 100 repeated measurements  over the five years mission, Gaia will hopefully achieve a positional error $p$ between $10$ to $200~\mu$as (for quasars
with magnitude $V=15$ to $20$). To compare our observations to Gaia
we need to evaluate the average $\Delta_{t}\gamma$ with $\Delta t=5$~years
and $N$ sources. The final Gaia error $p$ is obtained by best-fitting
$2N$ independent coordinates from $N^{2}/2$ angular separation measures;
the average positional error on the entire sky will
scale therefore as $(2N)^{-1/2}$. Over one hemisphere we can therefore
estimate that the error scales as $p/\sqrt{N}$. Since the average
angular separation of random points on a sphere is $\pi/2$, the average
of $\Delta_{t}\gamma$ can be estimated simply as $\Delta_{t}\gamma(\theta=\pi/2)$.
We find numerically $\Delta_{t}\gamma(\pi/2)=10\, s\,\mu$as, with
little dependence on $\Delta z$. Therefore Gaia can see the parallax
if $p/\sqrt{N}<10\, s\,\mu$as. For $s=4.4\cdot10^{-3}$ (i.e. the current
CMB limit) and $p=30\,\mu$as we need $N\gtrsim450,000$ sources: this shows
that Gaia can constrain the cosmic anisotropy to CMB levels. An enhanced
Gaia mission with $\Delta t=10$~years (or two missions 5~years
apart), $p=10\,\mu$as and $N=10^{6}$ would give $s<5\cdot10^{-4}$,
i.e. $r_{0}<1$~Mpc if we assume the sources are at 2~Gpc. Two caveats of this preliminary estimate are however to be kept in mind. First, to map the angular change, Gaia's observing strategy should be redesigned in order to maximize the time interval between quasar  observations. Second, being planned to monitor the local matter distribution, Gaia's errors estimates are based on using a fraction of the quasars as a reference frame. Observing the quasar CP would require a different statistical analysis. Both effects are likely to increase the expected final error.

Two local effects induce spurious parallaxes: one (of the order of
0.1$\mu$as$\, yr^{-1}$) is induced by our own peculiar velocity
and the other (of the order of 4$\mu$as$\, yr^{-1}$\cite{2003A&A...404..743K}
by a changing aberration. Both produce a dipolar signal, just like
a LTB: however, the peculiar velocity parallax decreases monotonically
with the angular diameter distance, while the aberration change is independent
of distance  \cite{2003A&A...404..743K}.
In contrast, the LTB signal has a characteristic non-trivial dependence
on redshift: for the models investigated here it is vanishingly small
inside the void, large near the edge, decreasing at large distances.
It is therefore possible in principle to subtract the cosmic signal
from the local one, for instance estimating the local effects from
sources inside the void, including Milky Way stars. A detailed calculation
needs a careful simulation of experimental settings (including possibly
effects like source photocenter jitter and relativistic light deflection
by solar system bodies) which is outside the scope of this paper.
Moreover, more general anisotropic models will not produce a simple
dipole.

\emph{Conclusions}.
Planned space-based astrometric missions aim at accuracies of the order of few microarcseconds. In this paper we have shown that the cosmic parallax of distant sources in an LTB model might be observable employing the same missions. Similar considerations would apply to all other anisotropic cosmological models as e.g. the Bianchi models. A positive detection of large-scale CP would disprove therefore one of the basic tenets of modern cosmology, isotropy. We have shown that for a typical LTB model designed to explain the supernovae Ia Hubble diagram a enhanced Gaia experiment could constrain the anisotropy parameter $s$ to less than $10^{-3}$, corresponding to the Megaparsec scale, much better than current CMB dipole limits \cite{Alnes:2006a}. Moreover, this test may probe a different range of scales depending on the quasar redshift distribution and, contrary to the CMB limits, the CP method cannot be completely undermined by the observer's peculiar motion and is limited only by source statistics instead of by the cosmic variance. We anticipate however that the major source of systematics would be the subtraction of the aberration change parallax. Real-time cosmology tests directly cosmic kinematics by observing changes in source positions and velocities. We have shown that the cosmic parallax, along with the velocity shift effect $\dot{z}$, can fully reconstruct the 3D cosmic flow of distant sources.

\emph{Acknowledgment.} We thank A. Balbi, R. Scaramella and S. Bonometto
for interesting discussions. MQ thanks Università di Milano-Bicocca
for support.

\vspace{-0.5cm}

\bibliography{Bltb}
 \bibliographystyle{apsrev}
\end{document}